# Confined States of a Positronium in a Spherical Quantum Dot

**Karen G. Dvoyan**

*Dept. of Physics and Technology, Russian-Armenian (Slavonic) University,*

*123 Hovsep Emin Str., Yerevan 0051, Armenia*

*Phone: (37493) 289-451; Fax: (374 10) 229-244; e-mail: dvoyan@gmail.com*

## Abstract

Confined states of an electron-positron pair in the spherical quantum dot (QD) are theoretically investigated in three size-quantization (SQ) regimes: strong, weak and intermediate. In the strong SQ regime, analytical expressions for the wave functions (WFs) and energy of the pair are obtained. In the weak SQ regime, the positronium energy and binding energy are analytically calculated. To calculate the positronium energy in the intermediate SQ regime, variational and numerical methods are used. It is shown that, in the corresponding limits, the results obtained by variational method agree with those obtained in the strong and weak SQ regimes.

## 1. Introduction

Recent interest to semiconductor QDs is conditioned by the new physical properties of these zero-dimensional objects, which are due to the SQ effect of the charge carriers (CCs) [1-7]. It is known that in the certain cases, when the low dimensional structure problems are considered, beside the SQ effects, the Coulomb interaction between CCs should be taken into account. It is also known that the dimensionality reduction of the sample leads to the amplification of the Coulomb interaction [8-11]. On the other hand, under certain conditions, the SQ effects can successfully compete with and prevail over the Coulomb quantization. For example, in the presence of the strong magnetic field, the Coulomb problem effectively becomes one-dimensional [8]. There is a huge number of theoretical and experimental works devoted to the investigation of the impurity and exciton states in QDs [12-16].

When solving the Coulomb problem in SQ systems, along with numerical methods, one has to apply various approximate methods. For example, the variational method is applied to find an electronic and impurity states in a spherical QD subject to homogeneous magnetic field [17]. At the same time, due to the significant difference of impurity (hole) and electron effective masses, one can apply the adiabatic approximation. It should also be mentioned that, in the cases when the SQ energy is much larger than the Coulomb one, the problem can be solved within the framework of the perturbation theory, by considering the Coulomb interaction as a small correction in the Hamiltonian of the problem [12].

The situation will radically change if one assumes that the effective mass of the impurity center is comparable with the electron mass. Such situations arise when considering, e.g., the Coulomb interaction of the electron-positron pair. As it is known, before the annihilation with electrons, the positrons decelerate in the condensed matter [18,19], thus forming a positronium both in the ground and excited states. There are two types of positronium: orto- (parallel orientation of spins) and parapositronium (antiparallel orientation of spins). Ortopositronium has a lifetime of $\tau \square \qquad {}^{7} s$, exceeding the parapositronium lifetime on three orders of magnitude. At annihilation, ortopositronium radiates three gamma quanta. It is known that the positronium origination cross-section is 43 times larger than the annihilation cross-section. Hence, it is expectable that in most cases before the annihilation the positronium can be formed.

A. P. Mills's group published a few articles devoted to the positronium creation on the internal pore surfaces implanted into a thin film of porous silica and positronium lifetimes measurements [20-22]. Wheeler supposed that two positronium atoms may combine to form the dipositronium molecule [23]. This molecule has been studied theoretically by Schrader [24]. Because positronium has a short lifetime and it is difficult to obtain low energy positrons in large numbers, dipositronium has not been observed unambiguously. A. P. Mills's group show that dipositronium is created on the internal pore surfaces when intense positron bursts are implanted into a thin film of porous silica. They found that molecule formation occurs much more efficiently than the competing process of spin exchange quenching. This result experimentally confirms the existence of the dipositronium molecule. As a purely leptonic, macroscopic quantum matter–antimatter system, this would be of interest in its own right, but it would also represent a milestone on the path to produce an annihilation gamma-ray laser [25]. The similarity between positronium and hydrogen atom allows for a rough estimate of energy levels $\varepsilon_p = \varepsilon_H / 2$, which are different between the two because of the positronium effective mass $\mu = m_e^* / 2$. Further, in the work [20] porous silica film contains interconnected pores with a diameter $d < 4nm$. From above mentioned follows, that it is logically necessary to discuss size quantization effects related with this topic.

As the foregoing theoretical investigation of positronium shows, the quantum states in the SQ semiconductor systems is a prospective problem of modern nanoscience. In particular, in the present paper, the confined states of a positronium in a spherical QD are theoretically investigated in three SQ regimes: strong, weak and intermediate.

## 2. Theory

Let us consider an impenetrable spherical QD. The potential energy of a particle in the spherical coordinates has the following form:

$$U(\rho,\theta,\varphi) = \begin{cases} 0, \rho \le R_0 \\ \infty, \rho > R_0 \end{cases}, \tag{1}$$

where $R_0$ is the radius of QD. The radius of QD and effective Bohr radius of the positronium $a_p$ play the role of the problem parameters, which radically affect the behavior of the particle inside QD. In what follows we analyze the problem in three SQ regimes: strong, weak and intermediate.

### The strong size quantization regime

In the regime of the strong SQ, when the condition $R_0 << a_p$ is satisfied, the energy of the electron-positron Coulomb interaction is much less, than the SQ energy. In this approximation, the Coulomb interaction between the electron and positron can be neglected, and the problem is reduced to the determination of the energies of the electron and positron, separately.

The Hamiltonian of the system in the spherical coordinates has the form

$$\hat{\mathrm{H}} = -\frac{\hbar^2}{2m_e^*}\nabla^2_{\vec{r}_e} - \frac{\hbar^2}{2m_p^*}\nabla^2_{\vec{r}_p} - \frac{e^2}{\kappa\left|\vec{r}_e - \vec{r}_p\right|}, \tag{2}$$

where $m_e^*$ and $m_p^*$ are effective masses of the electron and positron, respectively, $\kappa$ is the dielectric constant, $e$ is the particle charge.

The Hamiltonian of the system, when the Coulomb interaction is neglected, in the dimensionless quantities, can be written as

$$\hat{H} = -\frac{1}{2}\nabla^2_{\vec{r}_e} - \frac{1}{2}\nabla^2_{\vec{r}_p}. \tag{3}$$

Here $\hat{H} = \frac{\hat{\mathrm{H}}}{E_R}$, $r = \frac{\rho}{a_p}$, $\mu = \frac{m_e^* m_p^*}{m_e^* + m_p^*}$ is the effective mass of the positronium, $E_R^p = \frac{\hbar^2}{2\mu a_p^2} = \frac{\hbar^2}{m_e^* a_p^2} = \frac{e^2}{2\kappa a_p}$ is the effective Rydberg energy of the positronium, $a_p = \frac{\kappa\hbar^2}{\mu e^2} = \frac{\hbar^2}{m_p^* e^2}$ is the positronium effective Bohr radius. The WF of the problem can be sought in the form

$$\Psi\left(\vec{r}_e, \vec{r}_p\right) = \psi\left(\vec{r}_e\right)\Psi\left(\vec{r}_p\right), \tag{4}$$

where $\psi\left(\vec{r}_e\right)$ and $\Phi\left(\vec{r}_p\right)$ − are the WFs of the electron and positron, respectively. After some transformations, one can obtain the following expression for the WF:

$$\psi\left(\vec{r}_e\right) = \frac{1}{\sqrt{2\pi} r_0 J_{l+3/2}\left(\sqrt{2\varepsilon_r} r_0\right)} \frac{J_{l+1/2}\left(\sqrt{2\varepsilon_r} r\right)}{\sqrt{r}} Y_{lm}\left(\theta, \varphi\right), \tag{5}$$

where $J_{l+1/2}(z)$ are the first-kind Bessel functions of half-integer argument, $Y_{lm}(\theta, \varphi)$ are the spherical functions [26]. In this regime, one obtains the following expression for the electron and positron energies, respectively:

$$\varepsilon_e = \frac{\alpha_{n,l}^2}{2r_0^2}, \quad \varepsilon_p = \frac{\alpha_{n',l'}^2}{2r_0^2}, \tag{6}$$

where $r_0 = \dfrac{R_0}{a_p}$, $\alpha_{n,l}$ are the Bessel functions' roots, $n,l,m$ $(n',l',m')$ are the major, orbital and magnetic quantum numbers, correspondingly. The total energy of the system is the sum of the particle energies:

$$\varepsilon = \varepsilon_e + \varepsilon_p = \frac{\alpha_{n,l}^2 + \alpha_{n',l'}^2}{2r_0^2}. \tag{7}$$

**The weak size quantization regime**

In this case, when the condition $R_0 \gg a_r$ is fulfilled, the system's energy is mainly defined by the electron–positron Coulomb interaction. In other words, we consider the motion of a positronium as a whole in QD. The wave function of the system can be represented as

$$\Psi\left(\vec{r}_e,\vec{r}_p\right) = \psi\left(\vec{r}\right)\Phi\left(\vec{R}\right), \tag{8}$$

where $\vec{r} = \vec{r}_e - \vec{r}_p$, $\vec{R} = \dfrac{m_e^*\vec{r}_e + m_p^*\vec{r}_p}{m_e^* + m_p^*}$. Here $\psi\left(\vec{r}\right)$ describes the relative motion of the electron and positron, while $\Phi\left(\vec{R}\right)$ describes the positronium's center-of-mass motion. The Hamiltonian of the system is written as

$$H = -\frac{\hbar^2}{2M}\nabla_{\vec{R}}^2 - \frac{\hbar^2}{2\mu}\nabla_{\vec{r}}^2 - \frac{e^2}{\kappa\left|\vec{r}\right|}, \tag{9}$$

where $M = m_e^* + m_p^*$. In dimensionless quantities, this Hamiltonian takes the following form:

$$H = -\frac{1}{4}\nabla_{\vec{R}}^2 - \nabla_{\vec{r}}^2 - \frac{2}{r}. \tag{10}$$

One can obtain the energies of the center-of-mass motion and relative motion of the electron-positron pair, respectively:

$$\varepsilon_R = \frac{\alpha_{n,l}^2}{4r_0^2}, \qquad \varepsilon_r = -\frac{1}{N^2},\ N = 1,2,\ldots. \tag{11}$$

Finally, one can derive the following expression for the total energy:

$$\varepsilon_{ps} = \varepsilon_R + \varepsilon_r = \frac{\alpha_{n,l}^2}{4r_0^2} - \frac{1}{N^2}. \tag{12}$$

Defining the positronium binding energy as a difference of positron absence and presence energies in the QD, one can finally derive:

$$\varepsilon_{bind} = \varepsilon_e - \varepsilon_{ps} = \frac{\alpha_{n,l}^2}{4r_0^2} + \frac{1}{N^2}. \tag{13}$$

**The intermediate size quantization regime**

In this regime, when the condition $R_0$ □ $_r$ holds, the Coulomb interaction energy of the electron-positron pair is comparable with the SQ energy of QD walls. Note that in the intermediate SQ regime for the electron-positron pair considerably differs from that for electron-hole pair [12]. Thus, in the case of the electron-hole interaction, the electron's motion averages over the slow motion of the hole (adiabatic approximation) since the effective mass of the hole is much larger than that of the electron. In our case such averaging is impossible because of the equality of the electron and positron effective masses. Therefore, to solve the problem, we apply two approaches: variational method and numerical method.

*Variational method*

Let us represent the Hamiltonian (10) as a sum

$$\hat{H} = \hat{H}_1 + \hat{H}_2, \tag{14}$$

where

$$\hat{H}_1 = -\frac{1}{4}\nabla_{\vec{R}}^2, \quad \hat{H}_2 = \hat{H}_0 + \hat{H}', \quad \hat{H}_0 = -\nabla_{\vec{r}}^2, \quad \hat{H}' = -\frac{2}{r}. \tag{15}$$

The WF can be sought in the form (8). Repeating the calculation procedure for the system's center-of-mass WF, one can obtain

$$\Phi\left(\vec{R}\right) = \frac{1}{\sqrt{2\pi} r_0 J_{l+3/2}\left(\sqrt{4\varepsilon_R} r_0\right)} \frac{J_{l+1/2}\left(\sqrt{4\varepsilon_R} R\right)}{\sqrt{R}} Y_{lm}\left(\theta, \varphi\right). \tag{16}$$

For the center-of-mass motion energy (in this regime) one obtains (11), while, to determine the relative motion energy, the variational method should be applied [27,28]. The trial variational WF of the relative motion of the system can be sought in the form

$$\psi_v = C(\lambda)\psi_0 e^{-\lambda r}, \tag{17}$$

where $\lambda$ is a variational parameter, $C(\lambda)$ is the normalization factor, $\psi_0$ is the solution of the following Schrödinger equation:

$$\hat{H}_0\psi_0 = \varepsilon_0\psi_0 . \tag{18}$$

The solution of the equation (18) yields the following expressions for the energy and WF, accordingly:

$$\varepsilon_0 = \frac{\alpha_{n',l'}^2}{4r_0^2}, \tag{19}$$

$$\psi_0 = \frac{1}{2\sqrt{2\pi}r_0 J_{l'+3/2}\left(\sqrt{\varepsilon_0}2r_0\right)} \frac{J_{l'+1/2}\left(\sqrt{\varepsilon_0}r\right)}{\sqrt{r}} Y_{l'm'}(\theta,\varphi) . \tag{20}$$

Further, according to the variational method, we solve the Schrödinger equation with trial WF:

$$\left(\hat{H}_0 + \hat{H}'\right)\psi_v = \varepsilon_r(\lambda)\psi_v , \tag{21}$$

where $\varepsilon_r(\lambda)$ is the electron-positron pair relative motion energy. After calculation of the normalization constant, we obtain the following expression for the ground state WF:

$$\psi_0 = \frac{B(\lambda)}{r}\sin\left(\sqrt{\varepsilon_0}r\right)e^{-\lambda r}, \tag{22}$$

where

$$B^{-2}(\lambda) = 4\pi A, \tag{23}$$

$$A = \int_0^{2r_0} \sin^2\left(\sqrt{\varepsilon_0}r\right)\exp(-2\lambda r)\,dr \tag{24}$$

The variational  ground state energy can be found in the following form:

$$\varepsilon(\lambda) = \left\langle \psi_v^{(0)}(r)\left| -\left(\frac{\partial^2}{\partial r^2} + \frac{2}{r}\frac{\partial}{\partial r}\right) - \frac{2}{r} \right| \psi_v^{(0)}(r)\right\rangle = \varepsilon_0 - \lambda^2 - \frac{2B_1 + \lambda\sqrt{\varepsilon_0}B_2}{A}, \tag{25}$$

where the following notations have been introduced:

$$B_1 = \int_0^{2r_0} \frac{1}{r} \sin^2\left(\sqrt{\varepsilon_0} r\right) e^{-2\lambda r} dr,$$
$$B_2 = \int_0^{2r_0} \sin\left(2\sqrt{\varepsilon_0} r\right) e^{-2\lambda r} dr. \tag{26}$$

*Numerical method*

In this case, the second term of the Hamiltonian (14) can be represented as

$$\hat{H}_2 = -\nabla_{\vec{r}}^{2} \quad \frac{2}{r}. \tag{27}$$

Further, the problem is solved by using the method described for the weak SQ case, but the relative motion energy is defined by taking into account that the WF should be zero at QD walls.

**3. Discussion**

Now, we proceed to the discussion of the obtained results. As is can be seen, if the strong SQ regime is realized, the energy of the system is given as a sum of the electron and positron energies, which are quantized separately. In other words, in this SQ regime, the formation of the bound state of the pair is impossible. In Fig. 2 we plot the ground state energies of the electron and electron-positron pair as functions of the QD radius. Expectedly, due to the reduction of the confinement effects, increasing of the QD radius leads to the energy reduction. Note that the total energy is twice the single particle energy, because of the electron and positron masses equality.

In the weak SQ regime, due to the smallness of the SQ effects, the bound state of the electron-positron pair ─ positronium ─ is formed. In other words, the positronium center-of-mass motion is quantized in QD, and we have a situation analogous to exciton quantization in the weak SQ regime.

Figure 3 shows the dependence of the positronium's ground and the first excited state energies on the QD radius. This curve indicates that with the QD radius increasing, the Coulomb energy prevails over the SQ energy, and the energy becomes negative. In the limit of the large radii, the positronium energy converges to the free positronium energy value. An analogous behavior is observed for the first excited state, just one obtains four times larger value for the limit energy.

In Fig. 4 we plot the positronium binding energy and the hydrogen-like impurity energy as functions of the QD radius. As it can be seen from the figure, the decreasing of the QD radius leads to the increasing of the binding energy. However, expectedly, the binding energy of the hydrogen-like impurity is twice of the positronium binding energy. With increasing QD radius the positronium binding energy, the curve approaches the value of $E_{bind}^{p} = E_{R}^{p} = \frac{\hbar}{m_e^* a_p^2}$. Thus, for example, if the radius is equal to $R_0 = 2a_p$, one obtains the following value for the binding energy $E_{bind} = 1.62 E_R^p$, while at $R_0 = 5a_p$ the binding energy is $E_{bind} = 1.1 E_R^p$.

Let us proceed to the discussion of the electron-positron pair's behavior in the intermediate SQ regime, when the confinement and Coulomb interaction effects are of the same order. Figure 5 shows the dependence of the ground state energy of the electron-positron pair on the QD radius (in this regime), calculated by the two methods: variational and numerical. As it can be seen from the figure, the energy curves have analogous behavior in both cases. The energy difference of the variational and numerical cases is about $|\Delta E|$ [illegible] at $R_0 = 2a_p$, while $|\Delta E|$ [illegible] at the QD radius value $R_0 = 3a_p$. In other words, in the intermediate SQ regime, the electron-positron pair energy, calculated by applying the variational method, excellently matches the numerical result. It can be further seen that when increasing the QD radius, the discrepancy of the analytical and numerical results is reduced.

Finally, the dependence of the electron-positron pair ground state energy on the QD radius in all three SQ regimes is shown in Fig. 6. As it can be seen from this figure, the energy curve calculated by the variational method merges the strong SQ regime curve at small values of the QD radius. A similar behavior is also observed at large values of the QD radius. In this case, the variational curve almost coincides with the weak SQ regime curve. As it was expected, for the intermediate values of QD radius, a considerable discrepancy from the above-mentioned two regimes' curves is observed. As has been noted above, for the intermediate values of the QD radius ($R_0$ [illegible]), the intermediate SQ regime is realized, and the variational curve perfectly fits the numerical one (compare with Fig. 4).

## 4. Conclusion

In the present paper the confined states of the electron-positron pair in a spherical QD have been theoretically investigated in three SQ regimes: strong, weak and intermediate. Analytical expressions for the WFs and the energy of the pair have been obtained in the strong SQ regime, as well as the positronium and binding energies have been calculated in the weak SQ regime. The positronium energy has been calculated by using the variational and numerical methods in the intermediate SQ regime. It has been shown that in the limit of large radii the results obtained by using the variational method agree with those obtained for the strong and weak SQ regimes. At large QD radii, the convergence of the binding energy of the quantized positronium to the energy of the free positronium has been revealed. It has been shown that the results obtained by using the variational method converge to those obtained for the limiting cases of the strong and weak SQ. It has been shown that the positronium binding energy is half of the binding energy of the hydrogen-like impurity.

This theoretical investigation of the confined states of the positronium can be effectively used for the direct applications in photonics as the background for simulation model.

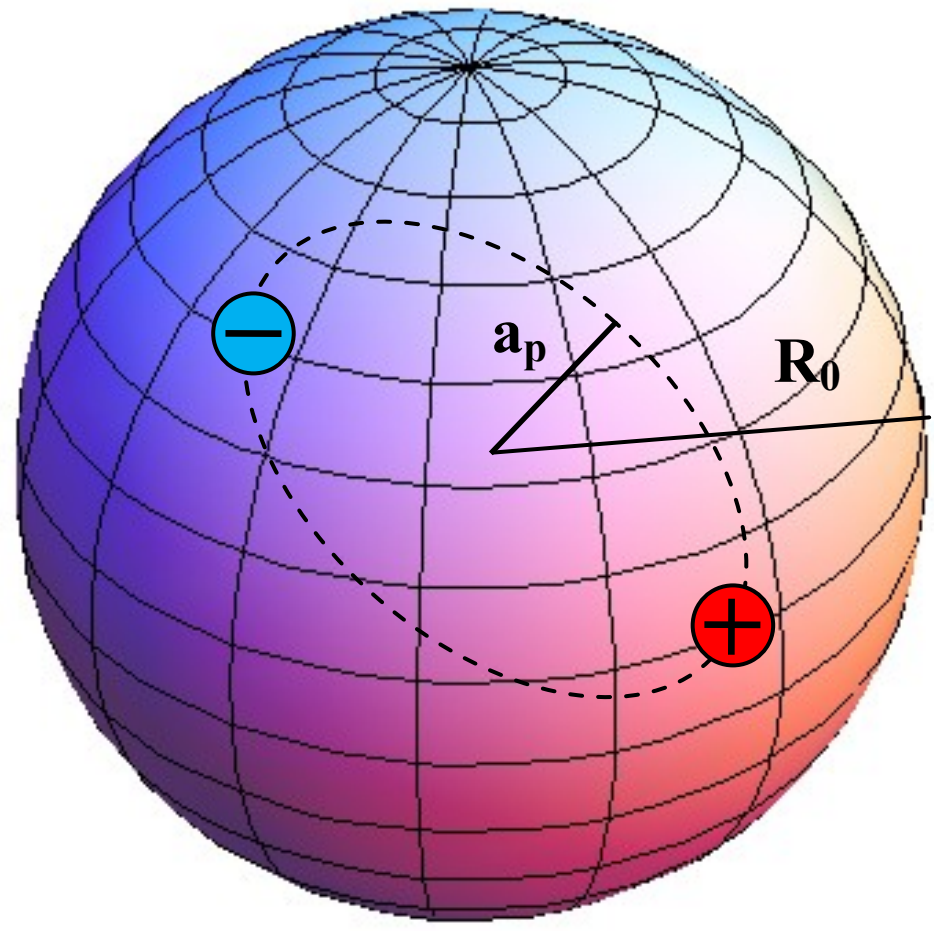

Fig. 1. Electron-positron pair in a spherical QD.

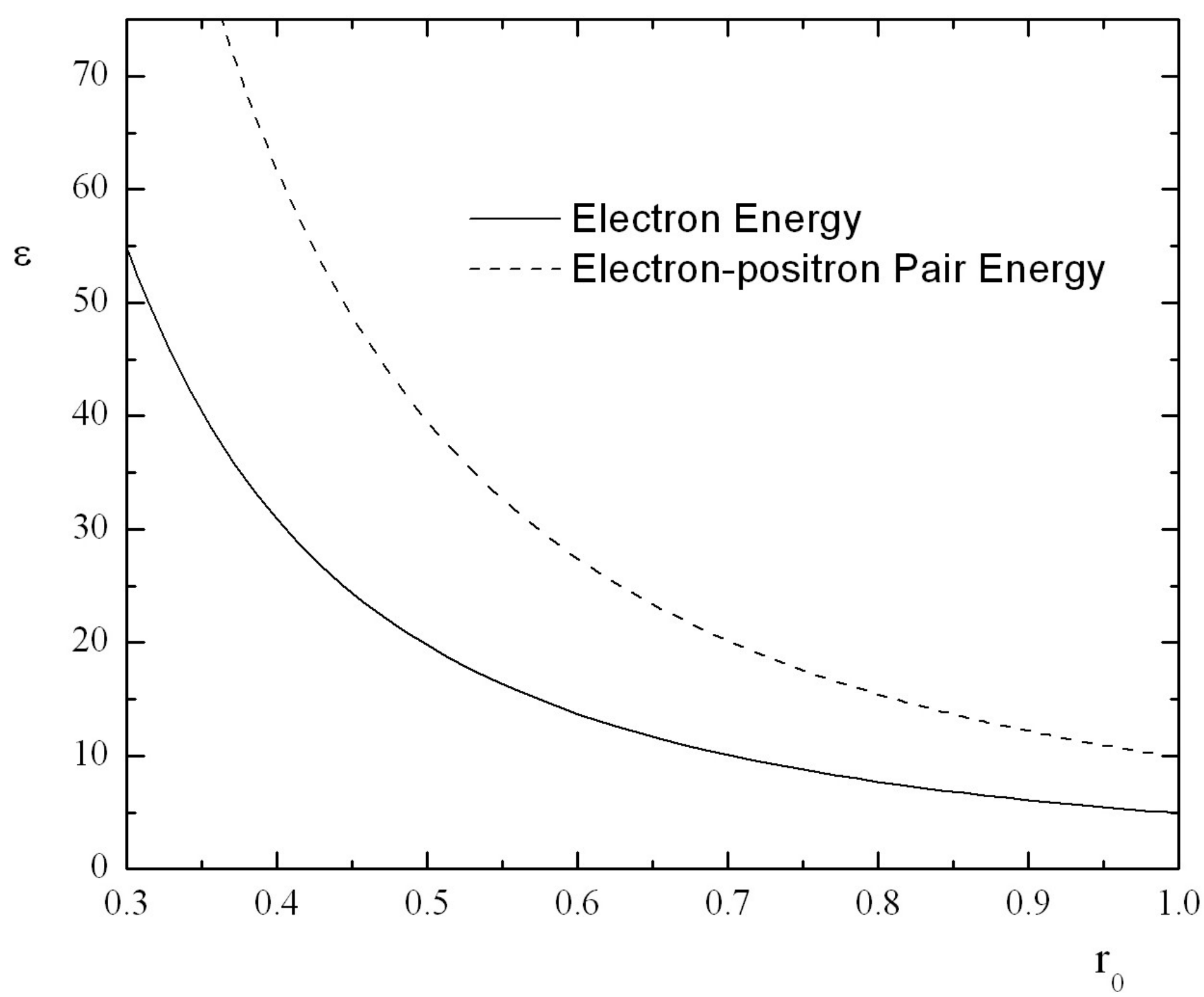

Fig. 2. Dependences of the electron ground state energy and the electron-positron pair energy on QD radius in strong SQ regime.

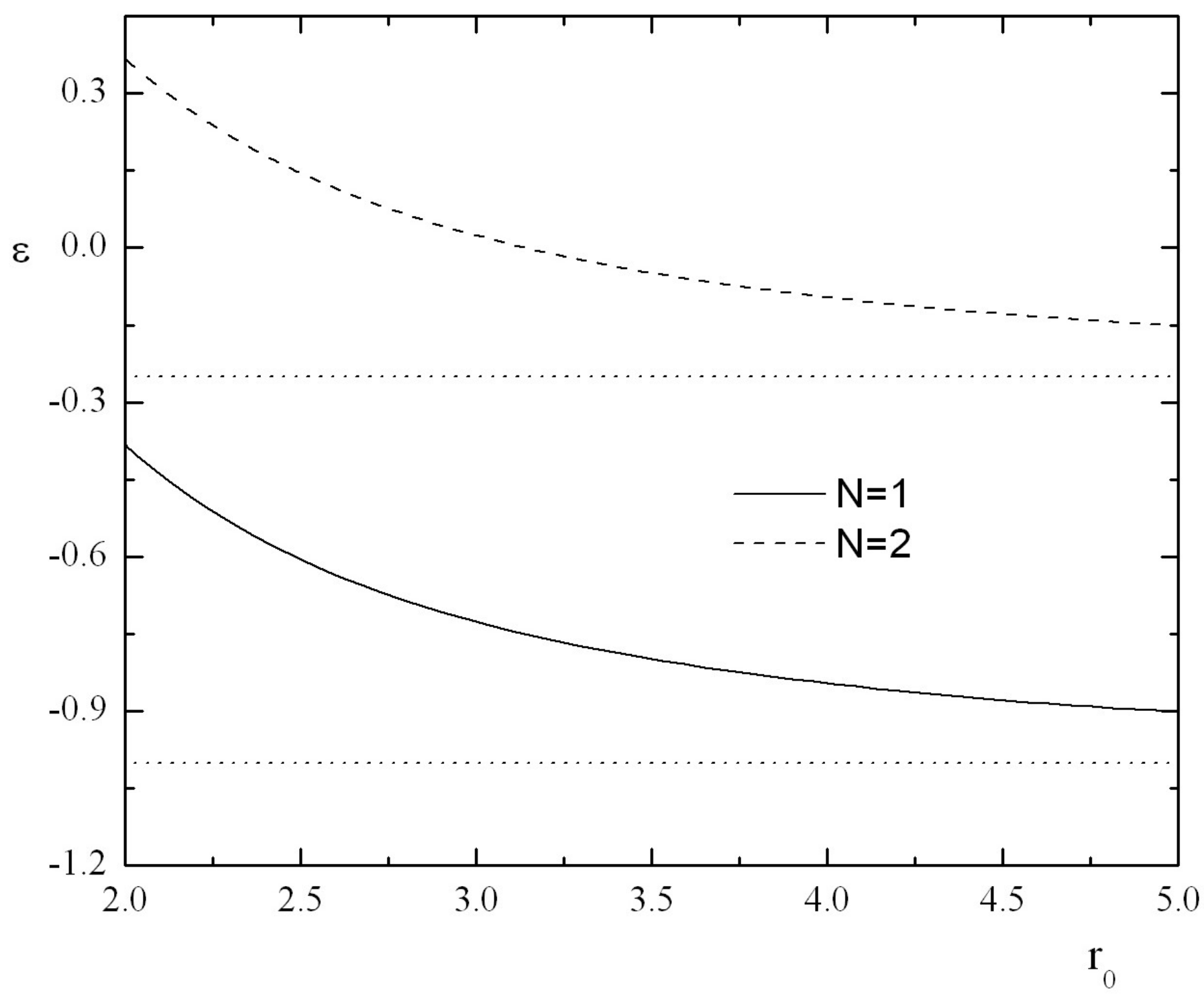

Fig. 3. Dependences of ground and the first excited state energies of positronium on QD radius in weak SQ regime.

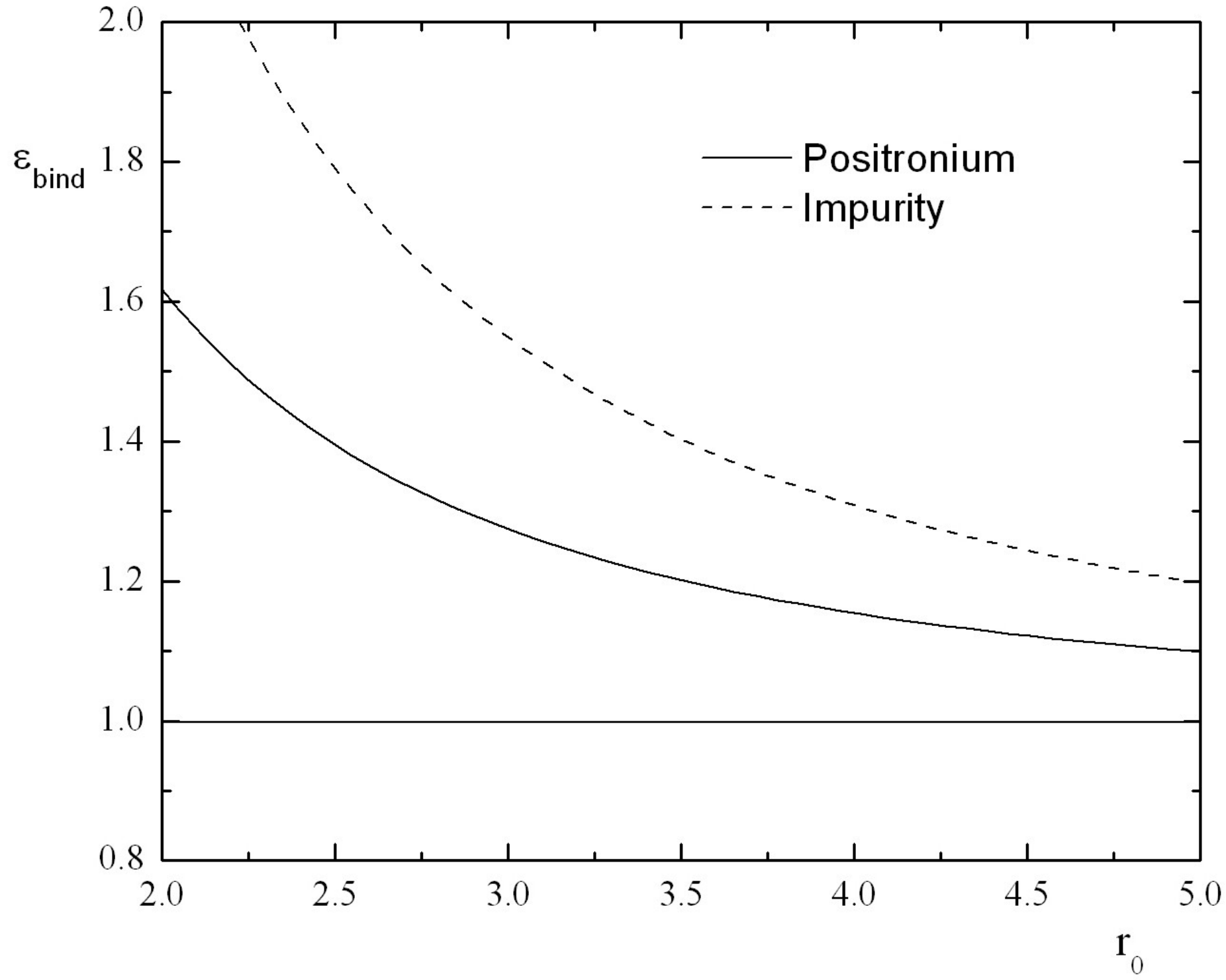

Fig. 4. Dependences of positronium and hydrogen-like impurity binding energies on QD radius.

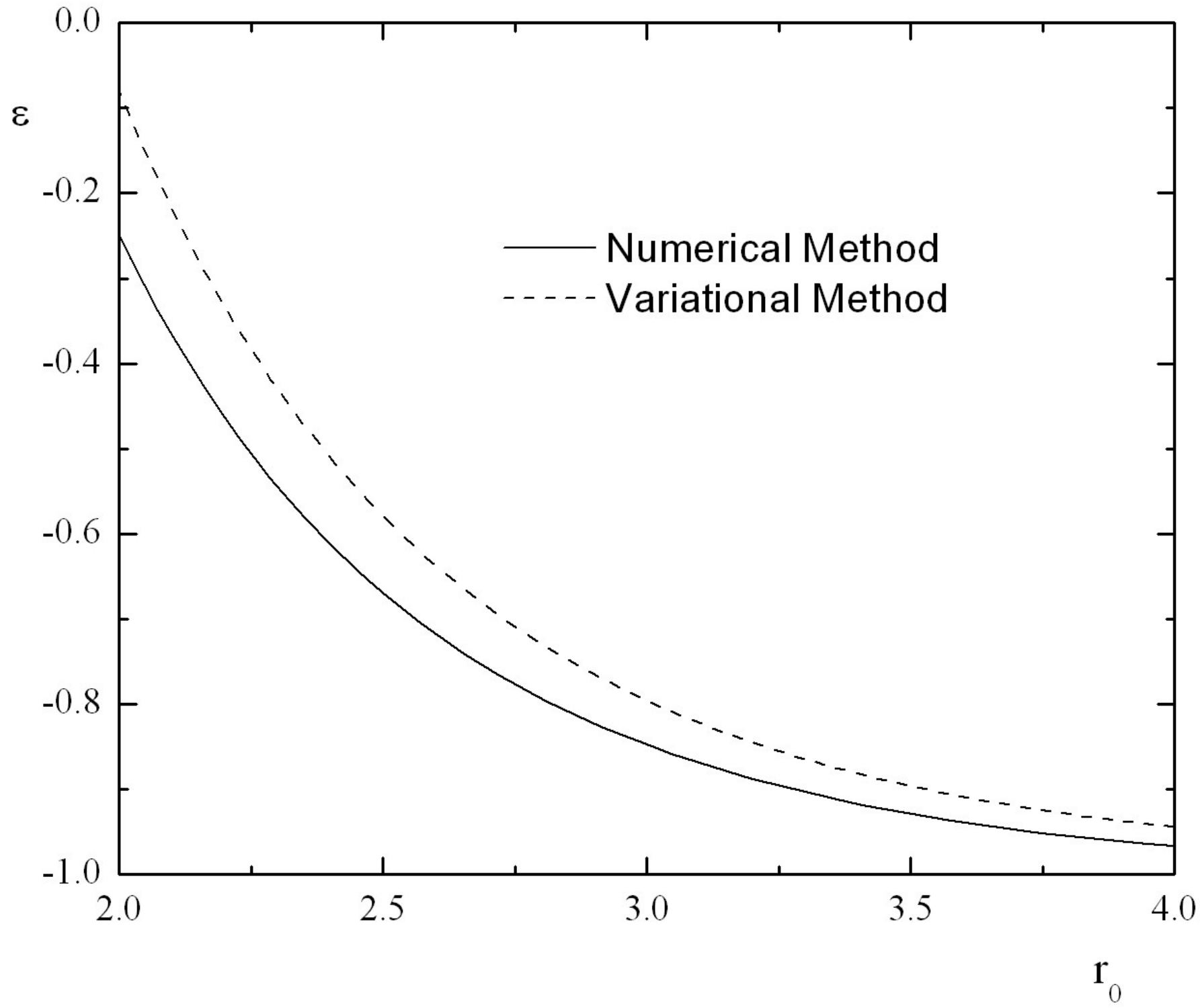

Fig. 5. Dependences of the electron and electron-positron pair ground state energies on QD radius in intermediate SQ regime.

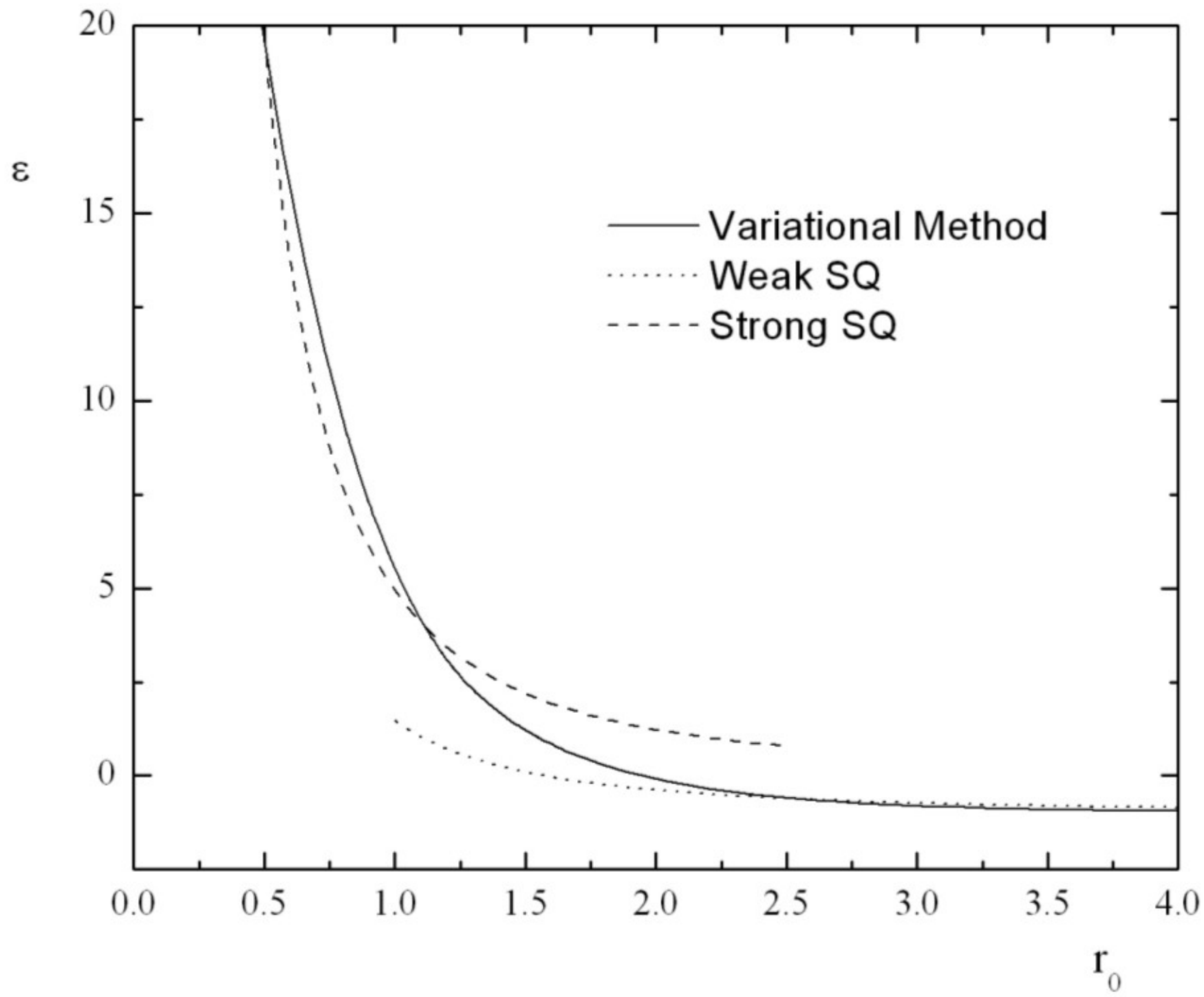

Fig. 6. Dependences of electron-positron pair ground state energies on QD radius in three SQ regimes.